\documentclass[showpacs,preprintnumbers,preprint]{revtex4}
\usepackage{graphicx,psfrag}
\usepackage{bm}

\begin{document}

%\preprint{APS/000}
\title{ Generation of strongly chaotic beats}

\author{I. \'{S}liwa}
\email{izasliwa@amu.edu.pl}
\author{P. Szlachetka}
\email{przems@amu.edu.pl}
\author{K. Grygiel}
\email{grygielk@amu.edu.pl}

\affiliation{Nonlinear Optics Division, Institute of Physics,
A. Mickiewicz University,
  Umultowska 85, PL 61-614 Pozna\'{n}, Poland }

\date{\today}

\begin{abstract}
The letter proposes a procedure for generation of strongly chaotic beats that have
 been hardly obtainable hitherto. The beats are
generated in a nonlinear optical system governing second-harmonic generation of light.
 The proposition is based on the concept of an optical coupler but
 can be easily adopted to other nonlinear systems  and  Chua's
 circuits.\\ \\
{\em Keywords}: Chaotic beats; optical coupler.
\end{abstract}
\pacs{05.45.Ac, 05.45.Gg, 42.65.Sf, 42.82.Et}

\maketitle

\section{Introduction}
In the last few years, an increase in the interest in the study of dynamical systems and design
 of nonlinear circuits generating chaotic beats has been observed
 [Grygiel \& Szlachetka, 2002; Cafagna \& Grassi, 2004-2006; \'{S}liwa {\em et. al.}, 2007].
Beats in a nonlinear oscillator appear if the oscillator becomes
quasi-periodic for any reason, for example, due to the frequency
detuning [Minorski, 1962]. Usually, in nonlinear systems,
 we observe repeated sequences of complicated revivals and collapses [Eberly {\em et. al.}, 1980;
 Abdalla {\em et. al.}, 2005].
 Chaotic revivals and collapses, in contradistinction
 to the quasi-periodic ones, do not have any   repeated structure.
 The degree of chaoticity of beats generated in a dynamical system is usually
confirmed with the Lyapunov exponents. This approach seems to be
necessary as intricate quasi-periodic beats frequently resemble
chaotic beats and {\em vice versa}. In particular, the numerical
procedure proposed by Wolf
  [Wolf {\em et. al.}, 1985] is a useful and efficient method to get such exponents
  (ordered from maximal to minimal value), known also as the spectrum of Lyapunov exponents.
 The first exponent is traditionally named the maximal Lyapunov exponent
(MLE), and its positive value is indicated of  a chaotic motion.
Two positive Lyapunov exponents denote hyperchaotic behaviour.
One  way to get chaotic beats from their quasi-periodic
counterpart  is to decrease  damping in the quasi-periodic
system. This action, usually, leads to generation of weakly
chaotic beats [\'{S}liwa {\em et. al.}, 2007]. Therefore, the
problem is to find a way of generation of strongly chaotic beats.
In this letter, we propose a solution to  this problem, by way of
example,
  in a simple nonlinear optical process -- second-harmonic generation of light,
using the concept of a typical optical coupler [Pe\'{r}ina, Jr. \& Pe\'{r}ina , 2000]
that is made of two identical systems interacting linearly with each other.  Let us first  consider the
generation of chaotic beats in a single system.

\section{Beats in single SHG-system}

We study the following dynamical system [Drummond {\em et. al.}, 1980; Mandel \& Erneux , 1982;
Gao, 2004]:
\begin{eqnarray}
\label{e1}
\frac{da}{dt}&=&-i\omega a -\gamma_{a} a +\epsilon a^{*}b +F e^{-i\Omega_{a} t}\,,\\
\label{e2}
\frac{db}{dt}&=&-i2\omega b-\gamma_{b}b  -0.5\epsilon a^{2}\,,
\end{eqnarray}
where the complex variables $a$ and $b$ represent the amplitudes of the fundamental
 and second-harmonics modes, respectively.  The parameter $\epsilon$ governs the
 nonlinear interactions
between both modes. The quantities $\omega$ and $2\omega$  are the
  frequencies of the fundamental and second-harmonic modes, respectively.
 The terms $\gamma_{a}a$ and $\gamma_{b}b$  describe the mechanism of loss.
 Moreover, the system is pumped by an external field
$F e^{-i\Omega_{a} t}$, where $F$  is an electric field amplitude at the frequency $\Omega_{a}$.
The parameters,  $\omega$, $\gamma_{a}$, $\gamma_{b}$, $\epsilon$, $F$, and  $\Omega_{a}$
 are taken to be real. If $\gamma_{a}>>\gamma_{b}$, the loss mechanism in the second harmonic
  mode is usually neglected, that is we assume $\gamma_{b}=0$. This simplification is
  frequently called a good frequency conversion limit.
The system (\ref{e1})--(\ref{e2}) has three periodic solutions. Namely,
 two coexisting solutions, when $\gamma_{b}=0$ and
$\Omega_{\pm}=\omega\pm 0.5F/\gamma_{a}$ \,\,[\'{S}liwa {\em et. al.}, 2007]:
\begin{eqnarray}
\label{e3}
a_{\pm}(t)&=&\frac{F}{\gamma_{a}} e^{-i\Omega_{\pm} t}\,,\\
\label{e4}
b_{\pm}(t)&=&\mp\frac{i}{2}\frac{F}{\gamma_{a}}e^{-i2\Omega_{\pm} t}\,,
\end{eqnarray}
and the resonance solution,
 when $\Omega=\omega$\,\, [Mandel \& Erneux , 1982]:
\begin{eqnarray}
\label{e5}
a(t)&=&(A+B) e^{-i\omega t}\,,\\
\label{e6}
b(t)&=&-\frac{\epsilon}{2\gamma_{b}} (A+B)^{2}  e^{-i2\omega t}\,,\\
\nonumber \label{e7} A&=&\sqrt[3]{
\frac{\gamma_{b}F}{\epsilon^{2}}+ \sqrt{\left
(\frac{2\gamma_{a}\gamma_{b}}{3\epsilon^{2}}\right)^{3}
 +\left(\frac{\gamma_{b}F}{\epsilon^{2}}\right)^{2}}}\,,\\
\nonumber \label{e8} B&=&\sqrt[3]{
\frac{\gamma_{b}F}{\epsilon^{2}}- \sqrt{\left
(\frac{2\gamma_{a}\gamma_{b}}{3\epsilon^{2}}\right)^{3}
+\left(\frac{\gamma_{b}F}{\epsilon^{2}}\right)^{2}}}\,.
\end{eqnarray}
We may expect that revivals and collapses (beats) appear in the
neighbourhood of the periodic solutions if the conditions of
periodicity  are not held. For  solutions (\ref{e3})--(\ref{e4})
the region of beat frequencies  is situated between the
frequencies $\Omega_{-}$ and $\Omega_{+}$ (see Fig.2 in
[\'{S}liwa {\em et. al.}, 2007]). Beats in this region are either
quasi-periodic or chaotic
 depending on the value of the damping constant $\gamma_{a}$.
Beats can also be generated  in the  neighbourhood of  solutions
(\ref{e5})--(\ref{e6}) if $\Omega_{a} \approx \omega$. However,
the generation of beats when the second harmonic mode is damped
($\gamma_{b}\neq 0$), is much less  efficient than in the good
frequency conversion limit, where $\gamma_{b}=0$. Let us shortly
consider this problem by way of a numerical example. The results
are presented in Fig.1.
%%%%%%%%%%%%%%%%%%%%%%%%%%%%%%%%%%%
%     FIG1   /   FIG2
%%%%%%%%%%%%%%%%%%%%%%%%%%%%%%%%%%%
The Lyapunov map shows the regions of  quasi-periodic beats,
chaotic beats and  purely periodic behaviour; individual colours
correspond to the values of the maximal Lyapunov exponents.
 The case of  beats generation corresponds  the inside of the parabola ,
while  the periodic states correspond to the region outside it.
   As seen, the regions corresponding to chaotic beats (positive value of MLE) are
 in the lower part of the parabola ( blue and yellow areas). MLE's are of the rank $10^{-3}$.
 Therefore, beats generated in the neighbourhood of the resonance $(\Omega_{a}=\omega)$
 are weakly chaotic. An attempt at enhancing the chaoticity of the beats by increasing
 the amplitude $F$ has proved ineffective. Assuming e.g. $F=20$ instead of $F=5$ the chaoticity
 increases but the structure of the beats is destroyed. Therefore,
the problem arises  how to force the system $(a,b)$ to generate
strongly chaotic beats in the presence of large damping that
allows only  generation of  periodic states or quasi-periodic
beats. In what follows, we propose a solution to this problem.

\section{Beats in SHG-coupler system}

Let us consider the system $(a,b)$ and its copy $(A,B)$ and join both systems linearly
in the following way:
\begin{eqnarray}
\label{s1}
\frac{da}{dt}&=&-i\omega a -\gamma_{a}a +\epsilon a^{*}b +Fe^{-i\Omega_{a}t}+s_{1}A
\,,\\
\label{s2}
\frac{db}{dt}&=&-i2\omega b  -\gamma_{b}b -0.5\epsilon a^{2}+s_{1}B
\,,\\
\label{s3}
\frac{dA}{dt}&=&-i\omega A -\gamma_{A}A +\epsilon A^{*}B +Fe^{-i\Omega_{A}t}-s_{2}a\,,\\
\label{s4}
\frac{dB}{dt}&=&-i2\omega B -\gamma_{B}B -0.5\epsilon A^{2}-s_{2}b\,.
\end{eqnarray}
In nonlinear optics, systems of this type are usually called
two-core couplers. Therefore, the subsystem $(a,b)$ interacts
with the subsystems $(A,B)$ {\em via} the $s_{1}$-terms and {\em
vice versa} $(A,B)$ interacts with  $(a,b)$ {\em via} the
$s_{2}$-terms. If, for example, $s_{2}$-terms are turned off then
the interaction is unidirectional and, consequently, $(a,b)$
plays the role only of a {\em receiver} whereas $(A,B)$ is simply
a {\em transmitter}. This type of interaction can be useful to
generate chaotic beats  within $(a,b)$  even if  the {\em free}
system $(a,b)$ is not able to generate (due to the large damping)
chaotic beats itself.
  Let us consider two simple examples.

\subsection{ Generation of chaotic beats in a periodic system forced by  external
quasi-periodic beats}

Let us first consider the dynamics of the system described by
equations (\ref{s1})-(\ref{s4}) if the interaction between the
subsystems  $(a,b)$ and $(A,B)$ is turned off ($s_{1}=s_{2}=0$).
Then, the  system $(a,b)$ for $\omega=10$, $\gamma_{a}=0.5$,
$\gamma_{b}=0.005$, $\epsilon=0.1$, $F=5$, $\Omega_{a}=10.5$ and
for the initial conditions $ a(0)=10$, $b(0)=-5i$, generates a
periodic state (or tends to this periodic state, as a steady
state, if it starts from  arbitrary initial conditions). The
periodic behaviour in $Re\,b(t)$-component is shown in Fig.2a.

The system $(A,B)$  for $\omega=10$, $\gamma_{A}=0.5$,
$\gamma_{B}=0.05$, $\epsilon=0.1$, $F=5$ and $\Omega_{A}=9.9$ and
for the initial conditions $ a(0)=10$ and $b(0)=-5i$, generates
quasi-periodic beats (the spectrum of Lyapunov exponents has the
form $\{-0.0396,-0.2203,-0.2214,-0.2346\}$). The beats are
presented in Fig.2b.

To get chaotic beats,  we now turn on the coupling in
(\ref{s1})-(\ref{s2}) unidirectionally , in such a way that the
subsystem $(a,b)$ plays the role of a {\em receiver}
($s_{1}\neq0$)  whereas $(A,B)$ is a {\em transmitter
}($s_{2}=0$). Then, the quasi-periodic beats are transmitted from
the subsystem $(A,B)$ into the periodic subsystem $(a,b)$ but not
{\em vice versa}. The coupling can be turned on at an arbitrarily
requested  time $t$. For $t=40$,
 $s_{1}=1.8$ and $s_{2}=0$  the  initially periodic system
 $(a,b)$ begins to generate  chaotic beats (Fig.3c).
 The choticity of the beats is confirmed by the spectrum of
Lyapunov exponents $\{0.0436,-0.1311,-0.4905,-0.8791\}$.
Therefore, the MLE ($\lambda_{1}=0.0257$) is two orders of
magnitude greater than its counterpart obtained in Section II.

The situation presented in Fig.2c changes rapidly if at the time
$t=40$, the feedback ($s_{2}\neq 0$)
 is turned on between the subsystems $(a,b)$ and $(A,B)$. Numerically, we have
put $s_{1}=s_{2}=1.8$. The new geometric structure of beats in
$Re\,b(t)$-component is  now presented in Fig.2d. As seen, after
$t>95$ the transient behaviour vanishes and finally we observe
beats. Also the structure of the beats in the subsystem $(A,B)$
changes due to the mutual interactions between the
  subsystems $(a,b)$ and $(A,B)$ . After $t>90$ we observe in
  $Re\,B(t)$-component repeated sequences of two
different beats (Fig.2e). It is obvious that beats in Figs.2d-e
are typically quasi-periodic, in contrast to those in Fig.2c. The
lack of chaoticity is confirmed by  the spectrum of Lyapunov
exponents $\{-0.1555,-0.1556,-0.1559,-0.1560\}$. Therefore, the
feedback interaction between  $(a,b)$ and
$(A,B)$ damages the chaos within beats in the subsystem $(a,b)$.\\
The degree of  stability of beats for different  values of the coupling parameter
 $s_{1}$ if $s_{2}=0$ is shown in Fig.3 (blue line).  Beats appear for $s_{1}>0.1$.
 For $0.1<s_{1}<1.43$ the beats are quasi-periodic, and
   in the range $1.43<s_{1}<5$ they behave chaotically. In this region
   the maximal Lyapunov exponent $\lambda_{1}$ increases exponentially with increasing
 value of the coupling parameter $s_{1}$, and generated beats become more and more chaotic.
 The numerical investigation shows that the system generates chaotic beats  since $s_{1}<10$.
 For $s_{1}>10$ the beats disappear at all but the system still remains chaotic.\\
It is worth noting that all the maximal Lyapunov exponents in
Fig.3 become negative if we turn on additionally the feedback
interaction ($s_{2}=s_{1}$) between the receiver and the
transmitter. Both subsystems still generate beats but they are
not chaotic.
%%%%%%%%%%%%%%%%%%%%%
%%%   FIG   3/4
%%%%%%%%%%%%%%%%%%%%%

\subsection{Generation of chaotic beats in a quasi-periodic system forced by other
quasi-periodic beats}

The periodic behavior presented in Fig.2a does not appear if, instead of the external
 frequency $\Omega_{a}=10.5$,
we put $\Omega_{a}=9.84$. Physically, it means that  the detuning  between the frequencies
$\omega=10$ and
$\Omega_{a}$ is now much  smaller. And consequently, in the subsystem $(a,b)$ instead of
the periodic behavior, beats appear (Fig.4a). Let us now suppose that the subsystem $(A,B)$
is in the same state as in  Subsection A (Fig.2b) and try to generate chaotic beats in the
 unidirectional regime $(A,B)\Rightarrow(a,b)$.  For $t=40$,
 $s_{1}=1.8$ and $s_{2}=0$ the receiver $(a,b)$ generates beats having complex geometrical
 structure but they are not chaotic which is confirmed by the Lyapunov exponents
   $\{-0.0221,-0.1550,-0.4972,-0.7829\}$. The coupling is simply too week to generate chaos
   within beats. However, a stronger coupling allow us to generate chaos.
For example, for $s_{1}=2.75$ and $s_{2}=0$ we observe distinctly
 chaotic beats in the subsystem $(a,b)$ which is confirmed by
the Lyapunov coefficients $\{0.0257,-0.0867,-0.5607,-0.8354\}$.
 As follows, the MLE $(\lambda_{1}=0.0257)$ is of the same order as that specified in Subsection A.
 The  geometric structure of beats in $Re\,b(t)$-component is
presented in Fig.4 .The degree of  chaoticity  of beats for different  values of the coupling parameter
 $s_{1}$ (if $s_{2}=0$)  in the range $0<s_{1}<5$ is illustrated in Fig.3 (green line).
 Quasi-periodic beats occur for $0<s_{1}<2.62$ and $2.81<s_{1}<2.92$.
 Chaotic beats are generated  in the range
 $2.63<s_{1}<2.81$ and  $2.97<s_{1}<5$. Similarly as in Subsection A, on turning
   on the feedback interaction ($s_{2}=s_{1}=2.75$) both subsystems generate
  beats  but they lose their  chaotic nature.

\section{Final remarks}
%We obtain  beats with higher degree of chaoticity than those  have been obtained till now.
We have proposed a procedure to obtain beats of a higher degree
of chaoticity than that hitherto obtained. Let us emphasize that
the method presented (unidirectional coupling between a receiver
and a transmiter) can be easily realized in arbitrarily dynamical
systems. The degree of chaos in the beats generated in  the
proposed optical transmiter-receiver system is of the same order
as in the R\"{o}ssler system [R\"{o}ssler, 1979]. The coupling
transmiter-receiver could be easily applied also in Chua's
circuits, where weakly chaotic beats have also been obtained
[Cafagna \& Grassi, 2004-2006].

%%%%%%%%%%%%%%%%%%%%%%%%%%%%
\newpage
\section{Figure Captions}

{\bf Figure 1.} The values of the maximal Lyapunov exponents
marked by separate colors for
 the system (\ref{e1})-(\ref{e2})
with $\omega=10$, $\epsilon=0.1$, $\gamma_{a}=0.5$, $F=5$,
$0<\gamma_{b}<0.2$ and $9.5<\Omega_{a} <10.5$.

{\bf Figure 2.} Time evolution of $Re\,b(t)$ and $Re\,B(t)$ for
the system (\ref{s1})-(\ref{s4}), where $\omega=10$,
$\gamma_{a}=0.5$, $\gamma_{b}=0.005$, $\Omega_{a}=10.5$,
$\gamma_{A}=0.5$, $\gamma_{B}=0.05$, $\Omega_{A}=9.9$,
$\epsilon=0.1$, $F=5$,
 and:\\(a)-(b) $s_{1}=s_{2}=0$;\\ (c)
$s_{1}=0$, $s_{2}=0$ if $t<40$ and $s_{1}=1.8$, $s_{2}=0$ if
$t>40$;\\(d)-(e) $s_{1}=s_{2}=0$ if $t<40$  and $s_{1}=s_{2}=1.8$
if $t>40$.\\The initial conditions are: $a(0)=10$, $b(0)=-5i$,
$A(0)=10$ and $B(0)=-5i$.

{\bf Figure 3.} The values of maximal Lyapunov exponents
$\lambda_{1}$, for the system (\ref{s1})-(\ref{s4}) as a function
of the coupling constant $s_{1}$, where $s_{2}=0$, $\omega=10$,
$\gamma_{a}=0.5$, $\gamma_{b}=0.005$, $\gamma_{A}=0.5$,
$\gamma_{B}=0.05$, $\epsilon=0.1$, $F=5$, $\Omega_{A}=9.9$ and
$\Omega_{a}=10.5$ (blue line), $\Omega_{a}=9.84$ (green line).

{\bf Figure 4.} Time evolution of $Re\,b(t)$ and $Re\,B(t)$ for
the system (\ref{s1})-(\ref{s4}), where $\omega=10$,
$\gamma_{a}=0.5$, $\gamma_{b}=0.005$, $\Omega_{a}=9.84$,
$\gamma_{A}=0.5$, $\gamma_{B}=0.05$, $\Omega_{A}=9.9$,
$\epsilon=0.1$, $F=5$,
 and:\\(a) $s_{1}=s_{2}=0$;\\ (b)
$s_{2}=0$, $s_{1}=0$ if $t<40$ and $s_{1}=2.75$, $s_{2}=0$ if $t>40$.\\
The initial conditions are: $a(0)=10$, $b(0)=-5i$, $A(0)=10$ and
$B(0)=-5i$.

%%%%%%%%%%%%%%%%%%%%%%%%%%

\newpage
%=================================
%            FIGURES             %
%=================================
%%%%%     FIG 1
\begin{figure}
\includegraphics[width=14cm,height=12cm,angle=0]{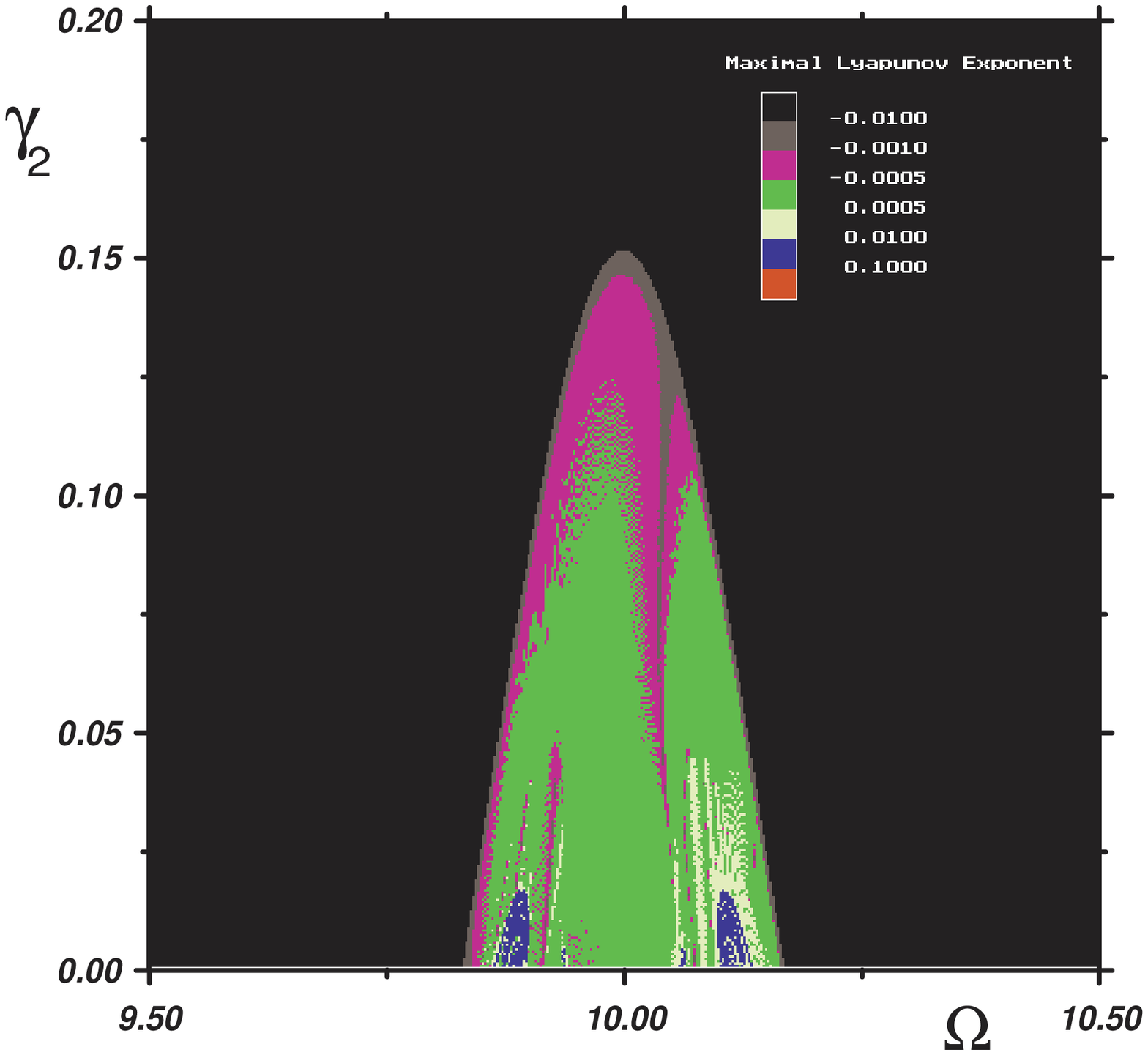}
\caption{ } \label{fig.1}
\end{figure}

%%%%%     FIG 2
\begin{figure}
\includegraphics[width=12cm,height=3.5cm,angle=0]{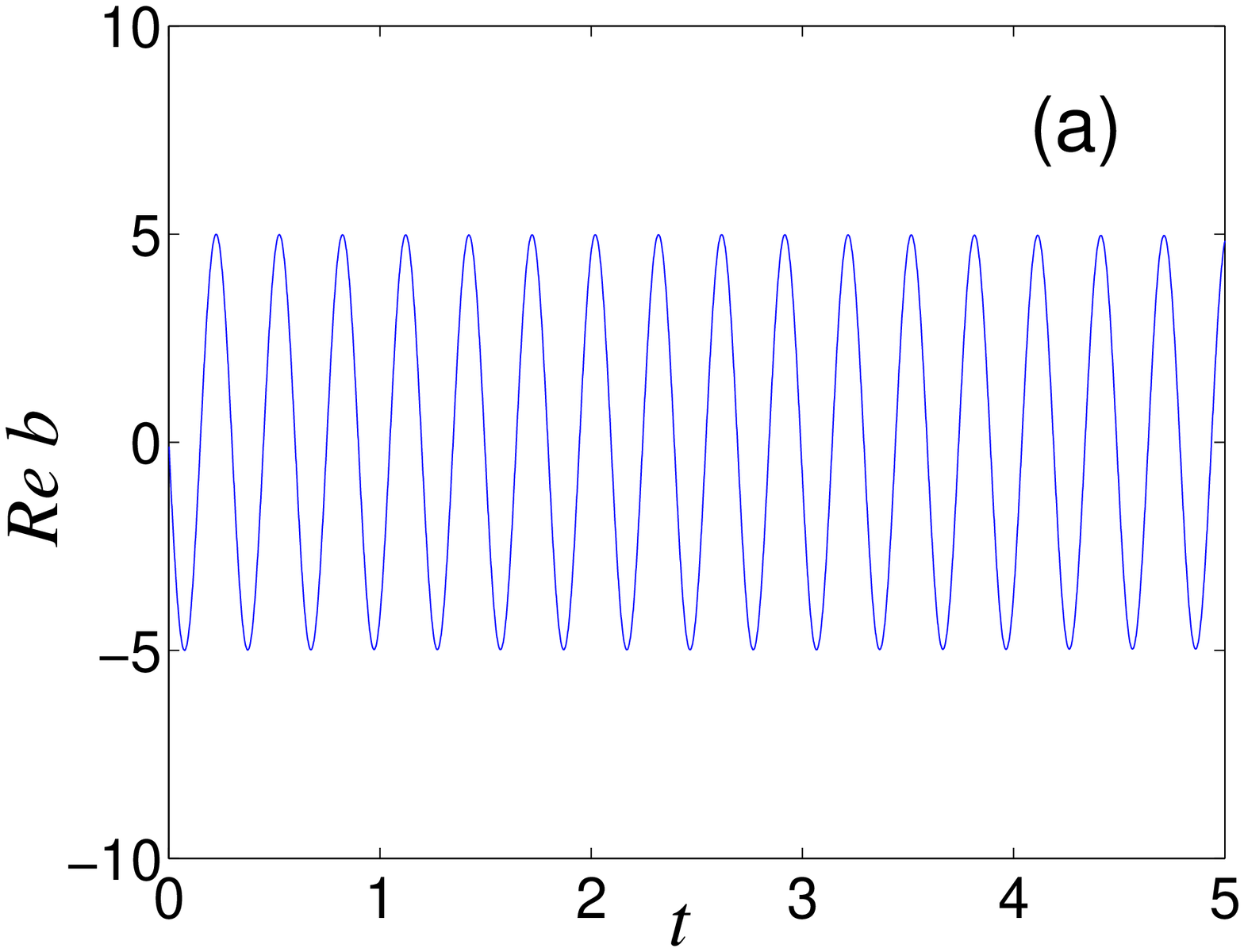}
\includegraphics[width=12cm,height=3.5cm,angle=0]{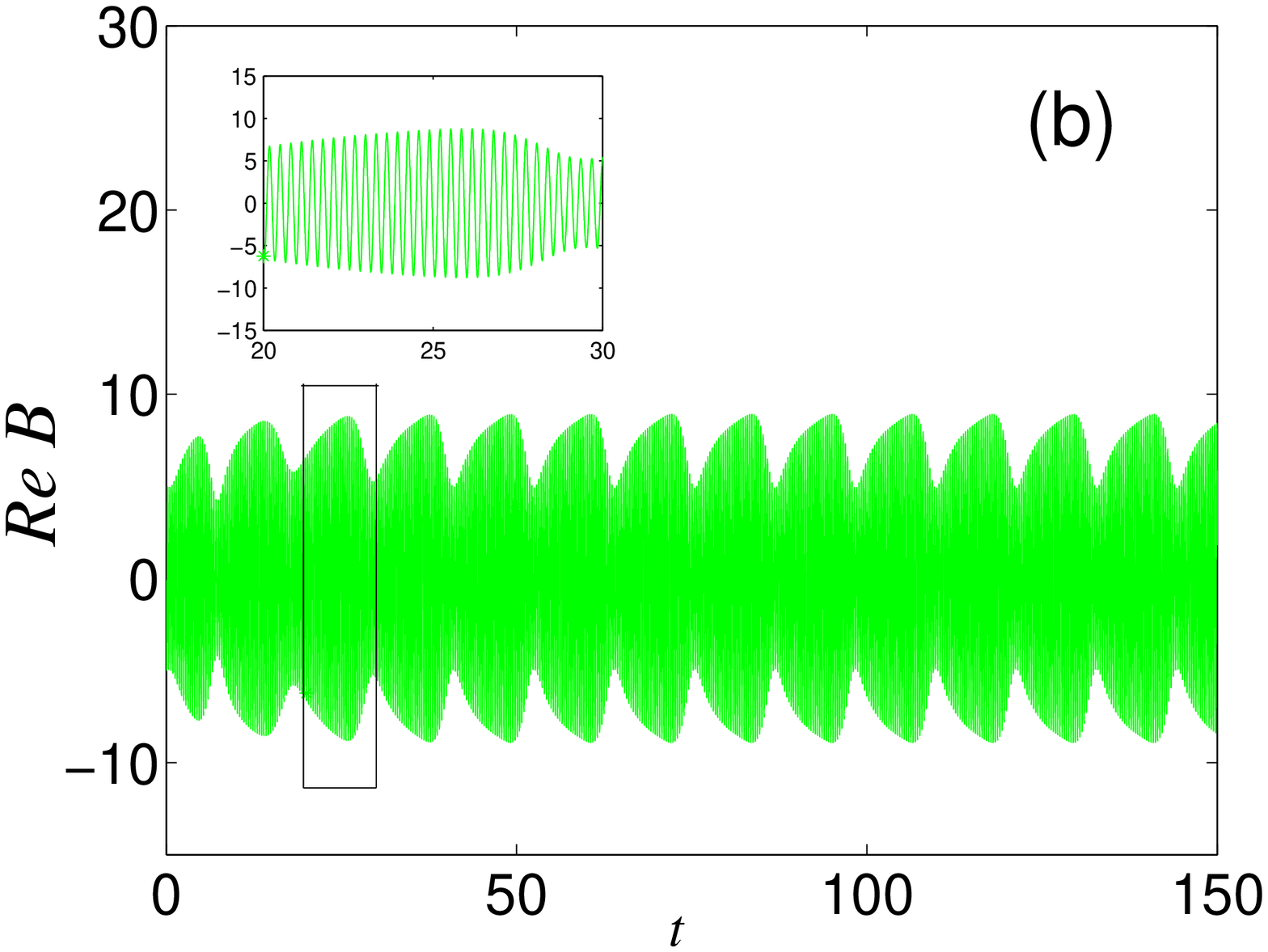}
\includegraphics[width=12cm,height=3.5cm,angle=0]{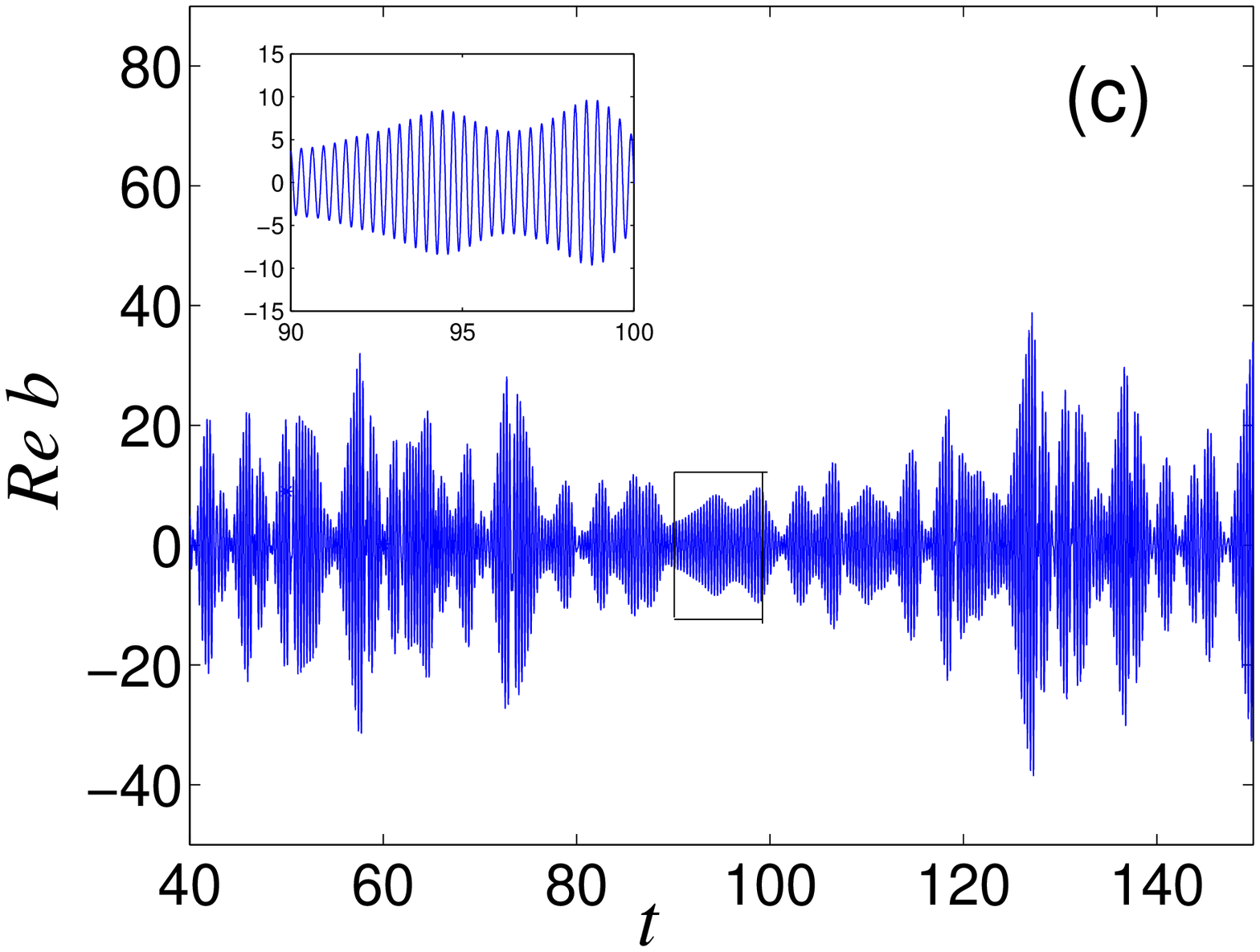}
\includegraphics[width=12cm,height=3.5cm,angle=0]{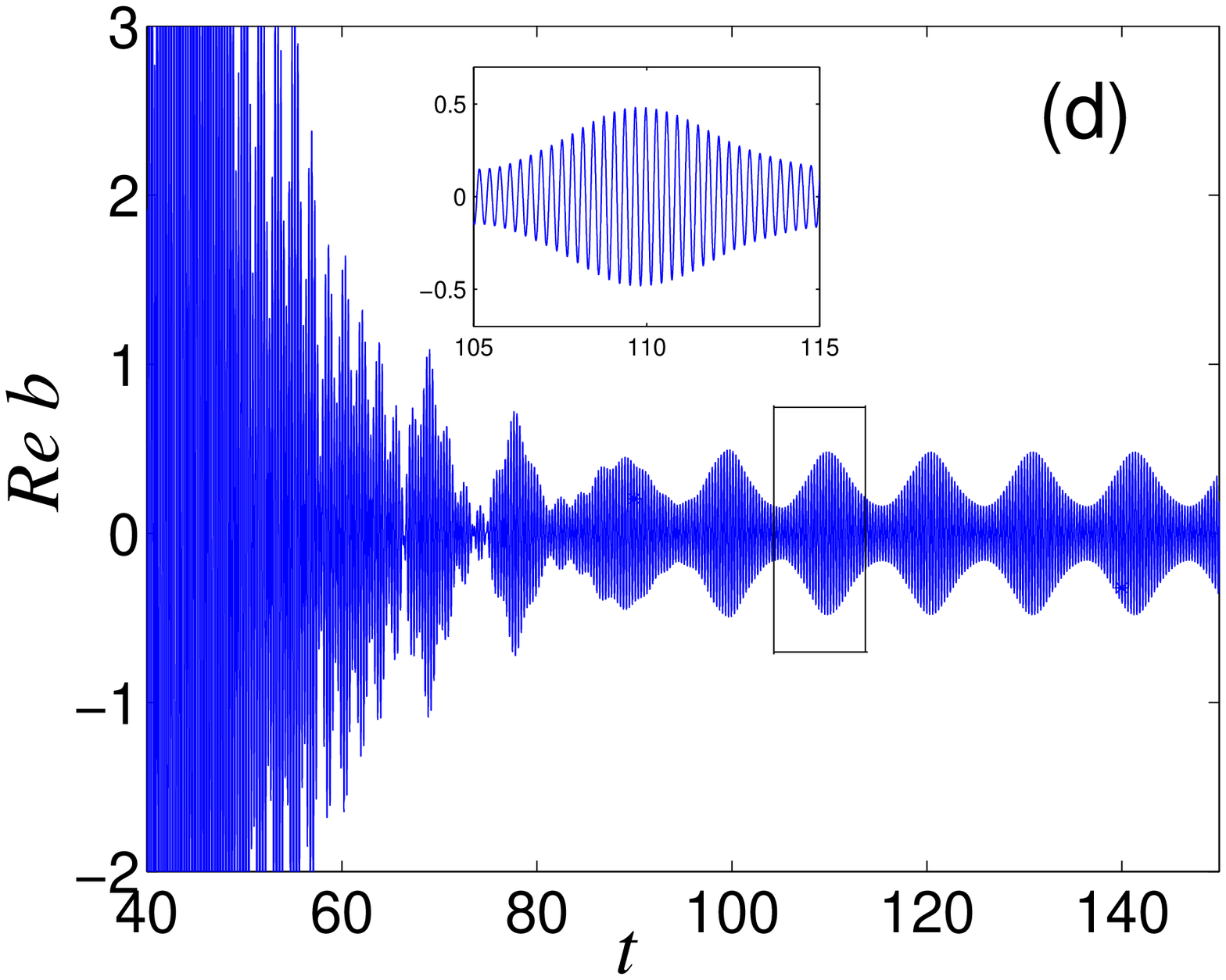}
\includegraphics[width=12cm,height=3.5cm,angle=0]{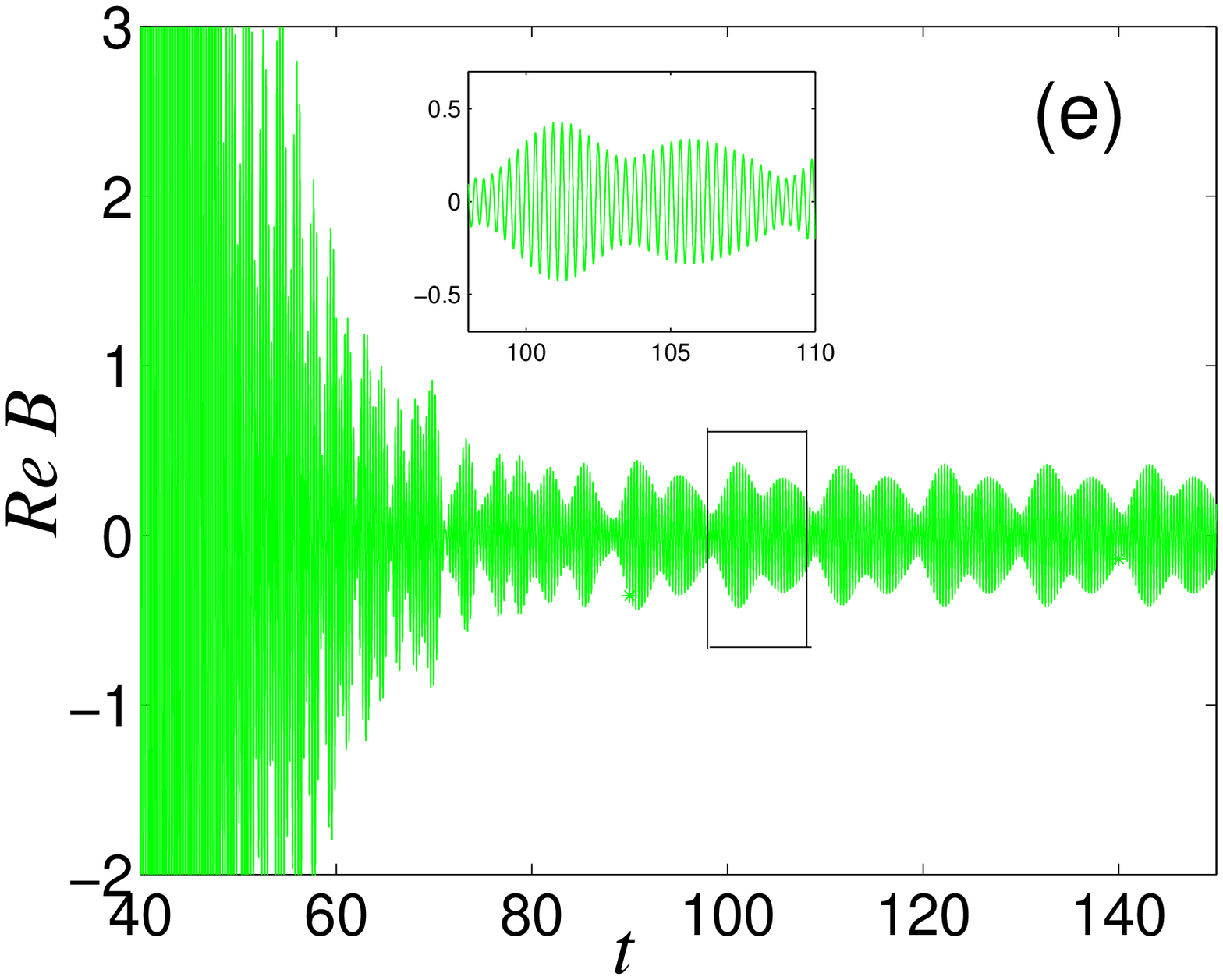}
\caption{ } \label{fig.2}
\end{figure}

%%%%%     FIG  3
\begin{figure}
\includegraphics[width=8cm,height=8cm,angle=0]{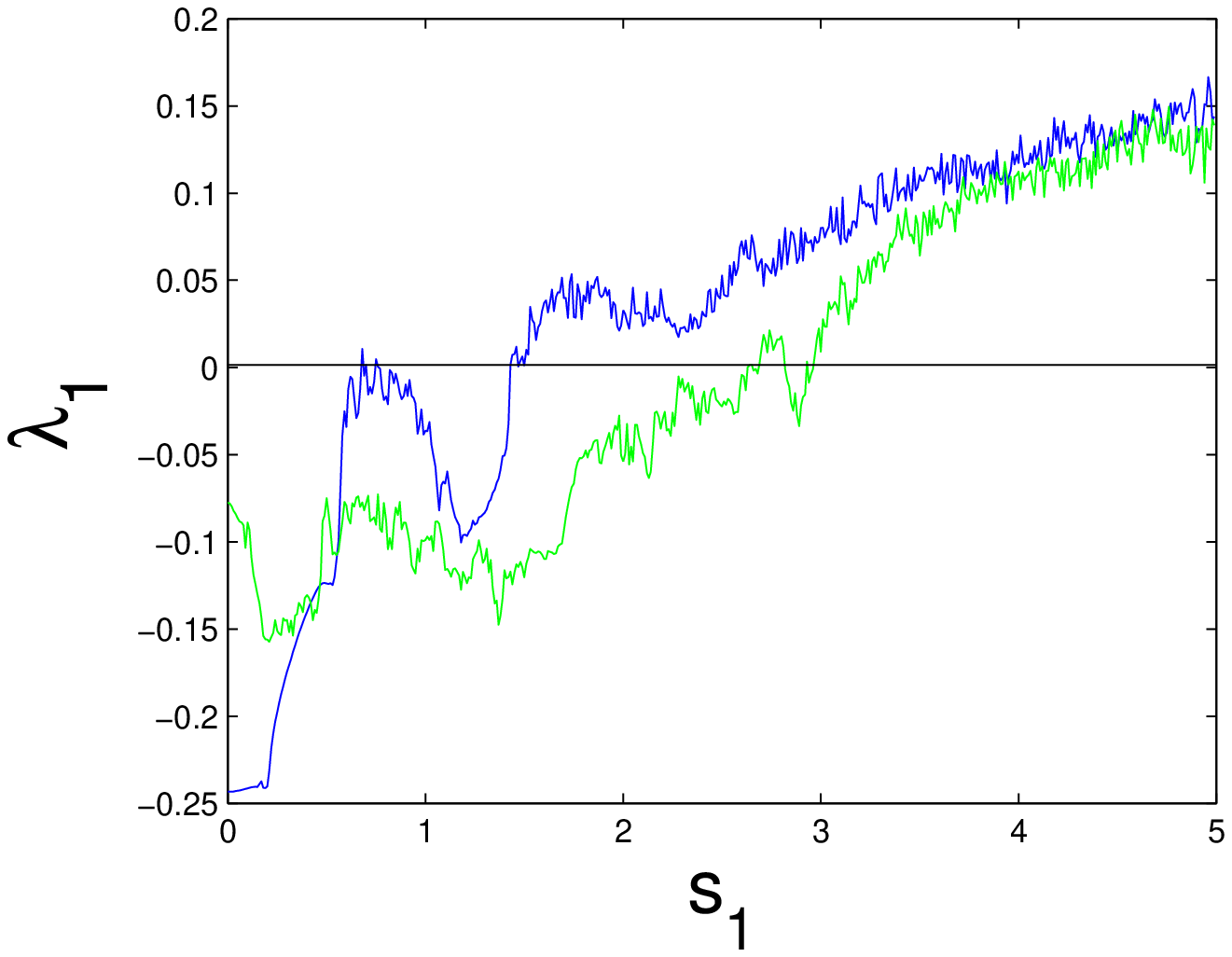}
\caption{ } \label{fig.3}
\end{figure}

%%%%%     FIG 4
\begin{figure}
\includegraphics[width=12cm,height=4cm,angle=0]{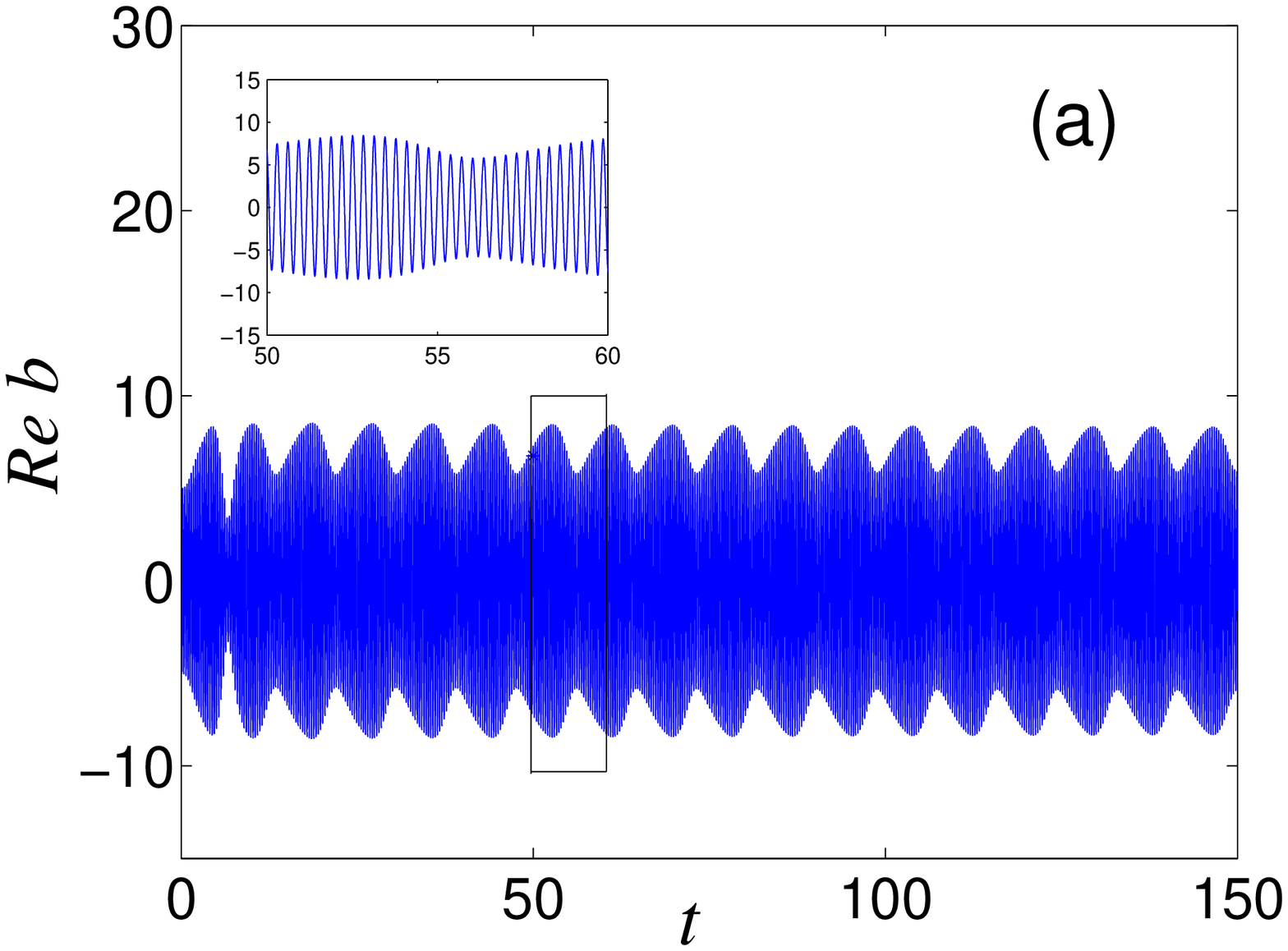}
\includegraphics[width=12cm,height=4cm,angle=0]{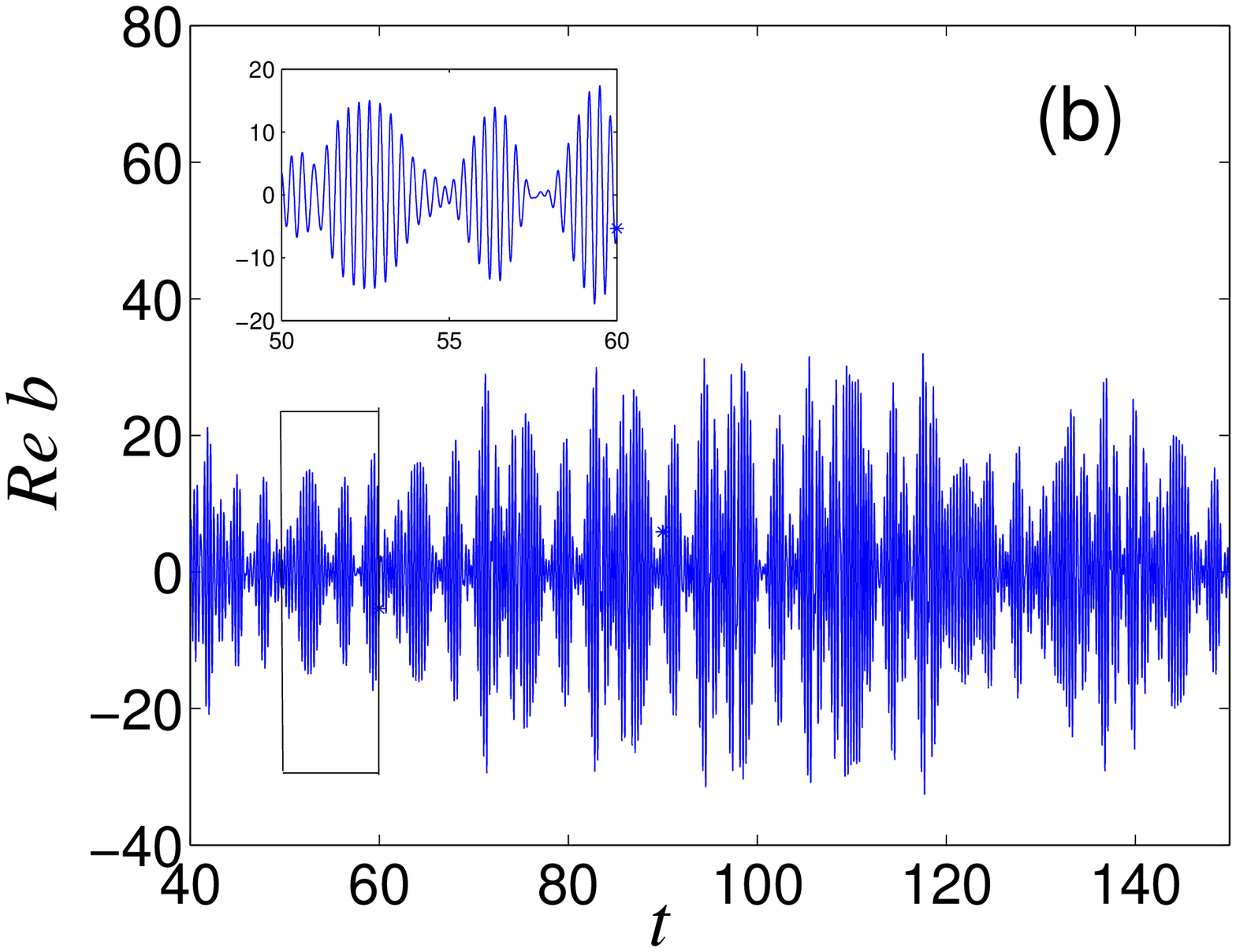}
\caption{ } \label{fig.4}
\end{figure}

%%%%%%%%%%%%%%%%%%%%%%%%%%%%%%%%%%%
\end{document}